\documentclass[manuscript]{aastex}
\usepackage{tipa}
\usepackage{amssymb}
\usepackage{txfonts}
\usepackage{graphicx}
\usepackage{cases}
\usepackage{booktabs,longtable}
\usepackage{lscape}
\usepackage[figuresright]{rotating}
\usepackage{multirow}
\usepackage{tabularx}
\usepackage{lineno}
\shorttitle{SSPM}
\shortauthors{Zhao et al}

\begin{document}


\title{COMPARISON OF CME/SHOCK PROPAGATION MODELS WITH HELIOSPHERIC IMAGING AND IN SITU OBSERVATIONS}


\author{XINHUA ZHAO\altaffilmark{1,2}, YING D. LIU\altaffilmark{1,3}, BERND INHESTER\altaffilmark{2}, XUESHANG FENG\altaffilmark{1}, THOMAS WIEGELMANN\altaffilmark{2}, AND LEI LU\altaffilmark{4,2,3}}
\altaffiltext{1}{State Key Laboratory of Space Weather, National
Space Science Center, Chinese Academy of Sciences, Beijing 100190,
China.} \altaffiltext{2}{Max-Planck-Institut f\"ur
Sonnensystemforschung, 37077 G\"ottingen, Lower Saxony, Germany.}
\altaffiltext{3}{University of Chinese Academy of Sciences, Beijing
100049, China.} \altaffiltext{4}{Key Laboratory of Dark Matter and
Space Astronomy, Purple Mountain Observatory, Chinese Academy of
Sciences, Nanjing, Jiangsu, China.}

\email{xhzhao@spaceweather.ac.cn;liuxying@spaceweather.ac.cn}


\begin{abstract}

The prediction of the arrival time for fast coronal mass ejections
(CMEs) and their associated shocks is highly desirable in space
weather studies. In this paper, we use two shock propagation models,
i.e. Data Guided Shock Time Of Arrival (DGSTOA) and Data Guided
Shock Propagation Model (DGSPM), to predict the kinematical
evolution of interplanetary shocks associated with fast CMEs. DGSTOA
is based on the similarity theory of shock waves in the solar wind
reference frame, and DGSPM on the non-similarity theory in the
stationary reference frame. The inputs are the kinematics of the CME
front at the maximum speed moment obtained from the geometric
triangulation method applied to STEREO imaging observations together
with the Harmonic Mean approximation. The outputs provide the
subsequent propagation of the associated shock. We apply these
models to the CMEs on 2012 January 19, January 23, and March 7. We
find that the shock models predict reasonably well the shock's
propagation after the impulsive acceleration. The shock's arrival
time and local propagation speed at Earth predicted by these models
are consistent with in situ measurements of WIND. We also employ the
Drag-Based Model (DBM) as a comparison, and find that it predicts a
steeper deceleration than the shock models after the rapid
deceleration phase. The predictions of DBM at 1 AU agree with the
following ICME or sheath structure, not the preceding shock. These
results demonstrate the applicability of the shock models used here
for future arrival time prediction of interplanetary shocks
associated with fast CMEs.
\end{abstract}


\keywords{ shock waves --- solar-terrestrial relations --- solar
wind --- Sun: coronal mass ejections (CMEs)}

\section{Introduction}
Coronal mass ejections (CMEs) are large-scale eruptions of plasma
and magnetic field from the Sun into interplanetary (IP) space. Soon
after their discovery in the 1970s, CMEs were regarded as major
sources for severe space weather events
\citep{Sheeleyetal1985,Gosling1993,Dryer1994}. For example, the
geoeffective CME would be a threat for astronauts performing space
activity, spacecraft, navigation \& communication systems,
airplane-passengers at high altitudes, ground power grids \&
pipelines (e.g., \cite{Boteleretal1998,Lanzerotti2005,NRC2008}).
Fast CMEs often lead to strong IP shocks ahead of them when
propagating in the heliosphere, and the shocks have additional space
weather effects, such as producing solar energetic particle (SEP)
events \citep{Gopalswamy2003,Cliver2009}, compressing the
geo-magnetosphere when colliding with the Earth
\citep{Greenandbaker2015}, and even causing a ``tsunami'' throughout
the whole heliosphere \citep{Intriligatoretal2015}. Therefore,
predicting arrival times of these fast CMEs/shocks at the Earth has
become a significant ingredient of space weather forecasting.
Various kinds of models in this aspect have been developed during
the past decades, such as empirical models, physics-based models,
and MHD models. \cite{zhaoanddryer2014} gave an overall review for
these models as well as their current applications.

The models in arrival time prediction usually adopt the observables
of CMEs/shocks obtained near the Sun as inputs to predict
whether/when they will arrive at the Earth. The in situ observations
at L1 spacecraft are then used to verify the prediction results. In
other words, the predictions are carried out only at two ends, i.e.
inputs on the Sun and outputs near the Earth. A prediction for the
CME/shock's propagation could not be checked by observations beyond
30 solar radii ($R_s$) in the heliosphere during the SOHO era
because the field of view (FOV) of SOHO/LASCO is within this
distance. The launch of STEREO in 2006 heralded a new epoch for
studies in this area. In contrast to SOHO, the FOV of the imaging
telescopes (HI1/HI2) onboard STEREO allows CMEs/shocks to be tracked
over much longer distances, even beyond the Earth's orbit.
Techniques have been developed to track solar disturbances based on
the wide-angle imaging observations of STEREO, e.g., the
triangulation technique developed by \cite{Liuetal2010a} has no free
parameters and can determine the CME/shock kinematics as a function
of distance. This triangulation technique initially assumes a
relatively compact CME structure simultaneously seen by the two
STEREO spacecraft. \cite{Lugazetal2010} and \cite{Liuetal2010b}
later incorporate into the triangulation concept to a spherical
front attached to the Sun for the geometry of wide CMEs (see
detailed discussions on the CME geometry assumptions in the
triangulation technique by \cite{Liuetal2010b,Liuetal2016}). The
triangulation technique with these two CME geometries has been
successfully applied to investigate the propagation of CMEs through
and their interactions with the inner heliosphere between the Sun
and Earth (e.g.,
\cite{Liuetal2010a,Liuetal2010b,Liuetal2013,Liuetal2016,Lugazetal2010,Daviesetal2013,
Mishra2013}).

In this paper, we present our improved shock propagation models that
utilize the early kinematics of the CME front as input, and predict
the shock's propagation in the subsequent IP space and its arrival
time at a given distance from the Sun. In particular, these
predictions will be verified not only by the in situ measurements at
1 AU but also by direct imaging of STEREO in IP space. This study
attempts to make qualitative comparisons between model predictions
and wide-angle imaging observations over long distances in the
heliosphere. The investigation will improve our understanding of the
kinematics of the CMEs/shocks during their outward propagation in IP
space.

\section{THEORETICAL MODELS}
\subsection{Data Guided Shock Time Of Arrival}
In the similarity theory of shocks, the similarity variable
$\eta=r/R(t)$ was adopted, where $r$ is distance and $R(t)$ is the
shock position at time $t$. Then a similarity solution was sought in
which the velocity, density, and pressure were rewritten as
functions of the non-dimensional similarity variable
\citep{Rogers1958,Sedov1959,Parker1961,Dryer1974}. Such a solution
maintains its similarity form. According to the similarity solution,
the shock's kinematics can be described in terms of normalized
ambient conditions, and are governed by its expansion effect. The
shock front keeps its shape when propagating outward. Once a shock
is identified in terms of a general limiting condition, then one may
note that it ``remembered'' its genesis \citep{Dryer1974}. In this
solution, the propagation speed ($V_b$) of a spherically symmetric
blast wave (no further energy ejected) in an ambient medium with
density decreasing in square of radial distance is
\citep{Parker1961,Dryer1974}:
\begin{eqnarray}\label{eq:sst}
V_{b}\sim R^{-0.5}
\end{eqnarray}
Here $V_b$ is in the reference frame of the ambient medium, $R$ is
the radial distance from the wave source. For the case of IP shock,
its propagation speed ($V_s$) in the stationary frame is computed as
$V_s=V_b+V_{SW}$; $V_{SW}$ is the ambient solar wind speed. As one
of the physics-based models in the shock arrival time prediction,
the ``Shock Time Of Arrival'' (STOA) model assumes that the shock is
initially driven by an ejecta at a constant speed. After this
initial driven phase, the shock front propagates outward as a blast
wave at the $R^{-0.5}$ deceleration speed
\citep{Dryer1974,dryerandsmart1984,smartandshea1984,smartandshea1985}.
The initial shock velocity is derived from the observed Type II
radio burst drifting speed, and the driving time duration is
estimated from the X-ray flux of the associated flare. Direct
observations of STEREO/SECCHI can provide the kinematics of the CME
front over long distances in the heliosphere, therefore we use
direct observations to replace the constant-speed assumption made in
STOA, named a Data Guided Shock Time Of Arrival (DGSTOA). In DGSTOA,
the $R^{-0.5}$ speed dependence relation is assumed to be valid for
the CME-associated shock after the CME completes its impulsive
acceleration and reaches the maximum speed near the Sun. Impulsive
accelerations of fast CMEs are found to be common
\citep{Zhangetal2001,Chengetal2010,Zhaoetal2010,Liuetal2011,Liuetal2013}.
Assuming that the shock has propagated to distance $R_M$ when the
front of the associated CME reaches its maximum speed $V_M$, then
the shock's propagation speed $V_s$ (in the stationary reference
frame) at any subsequent distance $R$ can be computed as:
\begin{eqnarray}\label{eq:vs_sstoa}
V_s=\frac{dR}{dt}=\left(V_M-V_{SW}\right)\left(\frac{R}{R_M}\right)^{-0.5}+V_{SW}
\end{eqnarray}
The numerical integral of $\frac{1}{V_s}$ provides the shock's
transit time $TT(R)$ from $R_M$ to $R$:
\begin{eqnarray}\label{eq:tr_sstoa}
TT(R)={\int_{R_M}}^R \frac{1}{V_s(R^\prime)}dR^\prime
\end{eqnarray}

In order to predict whether the shock can persist to the Earth, we
follow the judge index adopted in the STOA model. The ambient solar
wind and interplanetary magnetic field (IMF) parameters observed by
L1 spacecraft at the start time of the event are used to compute the
sound speed $C_S$ and  Alfv\'en speed $V_A$. Then the shock's
magnetoacoustic Mach number $M_a$ is derived as a measure of the
shock strength:
\begin{eqnarray}\label{eq:Ma}
M_a=\frac{V_{SE}-V_{SW}}{\sqrt{{C_S}^2+{V_A}^2}}
\end{eqnarray}
Here $V_{SE}$ is the shock's local speed at Earth computed from
Equation (\ref{eq:vs_sstoa}) (in the stationary reference frame).
The background solar wind speed $V_{SW}$ is taken from the in situ
measurements of L1 spacecraft at the start time of the event. If
$M_a
> 1$, the shock is predicted to reach the Earth; otherwise, it is
predicted to decay to become MHD waves before reaching the Earth.

\subsection{Data Guided Shock Propagation Model}

Different from the similarity theory, \cite{wei1982} and
\cite{weianddryer1991} studied in the stationary reference frame the
propagation of blast waves from a point source in a moving,
steady-state, medium with variable density. They considered both the
expansion effects of the shock and the convection effects of the
background flow. In their non-similarity solution, the shape and
kinematics of the shock are determined by the balance between the
initial blast energy and the background flow. They obtained the
following analytical solution to describe the propagation of the
blast wave:
\begin{eqnarray}\label{eq:vs_sspm}
V_s=\frac{dR}{dt}=\left[-2\lambda_1+\sqrt{(2\lambda_1)^2+\frac{E_0}{J_0R}+\frac{1}{2J_0}}\right]V_{SW}
\end{eqnarray}
Here, $V_s$ is shock speed in the stationary reference frame at
distance $R$, $t$ is time, $V_{SW}$ is solar wind speed (taken from
L1 spacecraft at the start time), and $E_0$ is the dimensionless
form of the initial total energy ($E_s$) that the blast released
into the background medium, i.e., $E_0=\frac{E_s}{AV_{SW}^2}$ where
$A$ = 300 kg~m$^{-1}$. $J_0=\frac{3}{8}$ and $\lambda_1=-0.1808$ are
constants of the coefficients of the solution. However, the energy
released by a blast cannot be observed directly.
\cite{fengandzhao2006} and \cite{zhaoandfeng2014} combined the
empirical method of estimating the shock's initial energy used in
the ISPM model \citep{smithanddryer1990,smithanddryer1995}, i.e.
$E_s=C.V_{si}^3.\omega.(\tau+D)$, with this non-similarity solution
of blast waves and proposed the ``Shock Propagation Model'' (SPM)
\citep{fengandzhao2006} and its second version (SPM2)
\citep{zhaoandfeng2014}. Here $V_{si}$ is the shock initial speed
computed from Type II drifting speed. $\tau$ is the piston-driving
time duration estimated from the X-ray flux of the associated flare.
$\omega$ denotes the solid angular width of the shock and is assumed
to be $60^\circ$ because it could not be determined accurately. $C$
(=0.283$\times10^{20}$ erg.m$^{-3}$.sec$^{-2}$.deg$^{-1}$) and $D$
(=0.52 hr) are two constants. \cite{zhaoandfeng2015} continued to
incorporate the kinematical parameters (velocity, angular width) of
CMEs observed by SOHO/LASCO with characteristics of the accompanied
solar flare-Type II events into the initial energy estimation of the
associated shock, and put forward a third version of the model
(SPM3). In SPM3, the prediction success rate on whether the shock
would arrive at Earth had been improved greatly. However, the
improvement in the arrival time prediction remained limited. The
reasons for this poor improvement mainly came from two aspects. On
one hand, the CME velocity used in \cite{zhaoandfeng2015} is the
projected speed on the plane of sky (POS) of SOHO, which does not
represent the propagation speed of the CME along the Sun-Earth
direction. On the other hand, the blast wave theory does not apply
to the initial acceleration phase of the CME/shock because there is
no additional energy added to the system in the blast wave theory.
The impulsive acceleration phase demonstrates that continuous energy
is still energy fed into the CME/shock system which leads to the
impulsive acceleration. Therefore, we decided to present a Data
Guided Shock Propagation Model (DGSPM) that uses the tracking
results of STEREO for fast CMEs during their initial acceleration
phases as inputs to predict the propagation and arrival time of
their associated shocks in the following deceleration phases.

Let us assume that the eruption occurs at $t=0$. The CME front
reaches its maximum speed $V_M$ when $t=t_M$ at the radial distance
$R_M$. The DGSPM model assumes that the blast wave theory starts to
be valid for the CME-associated shock at $R_M$. According to
Equation (\ref{eq:vs_sspm}):
\begin{eqnarray}\label{eq:vm_sspm}
V_M=\left[-2\lambda_1+\sqrt{(2\lambda_1)^2+\frac{E_0}{J_0R_M}+\frac{1}{2J_0}}\right]V_{SW}
\end{eqnarray}
Therefore,
\begin{eqnarray}\label{eq:e0_sspm}
E_0=J_0R_M\left[\left(2\lambda_1+\frac{V_M}{V_{SW}}\right)^2-(2\lambda_1)^2-\frac{1}{2J_0}\right]
\end{eqnarray}
From this equation we can estimate the initial energy of a shock
given $V_M$ and $R_M$ of the associated CME front. Then, the
integral of $\frac{1}{V_s}$ from Equation (\ref{eq:vs_sspm}) gives
the transit time $TT(R)$ of the shock to any subsequent distance
$R$:
\begin{eqnarray}\label{eq:tr_sspm}
TT(R)=\frac{J_0}{V_{SW}}\left\{4\lambda_1\left[R+2E_0-2E_0ln(R+2E_0)\right]+2\sqrt{\frac{E_0}{J_0}R+\left(4\lambda_1^2+\frac{1}{2J_0}\right)R^2}-\frac{\left(16\lambda_1^2+\frac{1}{J_0}\right)E_0}{\sqrt{4\lambda_1^2+\frac{1}{2J_0}}}\nonumber \right. \\
\times
ln\left[\sqrt{\frac{E_0}{J_0}R+\left(4\lambda_1^2+\frac{1}{2J_0}\right)R^2}+(R+2E_0)\sqrt{4\lambda_1^2+\frac{1}{2J_0}}-\frac{\left(16\lambda_1^2+\frac{1}{J_0}\right)E_0}{2\sqrt{4\lambda_1^2+\frac{1}{2J_0}}}\right]\nonumber\\
\left.-8\lambda_1E_0ln\left[\frac{\sqrt{\frac{E_0}{J_0}R+\left(4\lambda_1^2+\frac{1}{2J_0}\right)R^2}+4\lambda_1E_0}{R+2E_0}-\frac{16\lambda_1^2+\frac{1}{J_0}}{8\lambda_1}\right]\right\}+TT_0~~~~~~~~~~~~
\end{eqnarray}

Here, $TT_0$ is determined by the restriction of $R=R_M$ when
$TT=t_M$. Accordingly, the propagation speed of the shock at
distance $R$ is again computed from Equation (\ref{eq:vs_sspm}).

In order to predict whether or not the shock will persist to the
Earth, an index suggested by \cite{zhaoandfeng2015} is adopted in
the DGSPM. The Equivalent Shock Strength Index (ESSI) is defined as:
\begin{eqnarray}\label{eq:essi}
ESSI=\frac{V_s(EL)-V_{SW}}{V_F}
\end{eqnarray}
Here, $V_s(EL)$ is the local speed of shock at Earth, $V_F$ is the
fast-mode wave speed of the background solar wind at Earth, given by
$V_F^2=0.5[V_A^2+C_S^2+\sqrt{(V_A^2+C_S^2)^2-4V_A^2
C_S^2cos^2\theta}]$. $\theta$ is the angle between the shock normal
and the local magnetic field. In this study, we use the ambient
solar wind and IMF conditions in situ observed by the L1 spacecraft
at the start time of the event to compute $V_A$, $C_S$, and $V_F$
($\theta$ assumed to be 45$^\circ$ or 135$^\circ$). A threshold
value (ESSI$_{tv}$) needs to be set to predict whether the shock
could persist to the Earth. \cite{zhaoandfeng2015} tested 498 events
of Solar Cycle 23 and found that ESSI combined with the CME angular
width ($AW_{CME}$) could give the maximum successful predictions.
Here, $AW_{CME}$ is the angular width of the CME projected onto the
POS of SOHO. \cite{zhaoandfeng2015} set the threshold values as
ESSI$_{tv} = 1.53 $ and $AW_{tv} = 121^{\circ}$ by fitting all the
498 events. As a study for case events, we do not need to set new
threshold values. We just follow the criteria established in
\cite{zhaoandfeng2015}, i.e.: if ESSI $\geqslant$ ESSI$_{tv} = 1.53
$ and $AW_{CME} \geqslant AW_{tv} = 121^{\circ}$, then the shock is
predicted to be able to persist to the Earth, and its transit time
is predicted based on Equation (\ref{eq:tr_sspm}); Otherwise, the
shock is predicted to miss the Earth.

\section{APPLICATION TO THREE EVENTS}
Based on the wide-angle imaging observations of COR2, HI1, and HI2
onboard two STEREO spacecraft, \cite{Liuetal2013} studied the whole
Sun-to-Earth propagation of three CME events during 2012, which are
2012 January 19 (Case 1), 2012 January 23 (Case 2), and 2012 March 7
CMEs (Case 3). They are fast CMEs with maximum initial speeds of
1300--2300 km/s. These fast CMEs are not very common during the
long-lasting low solar active period, and are interesting subjects
to the space weather community. The tracking for these CME fronts
was implemented by a combination of the geometric triangulation
technique developed by \cite{Liuetal2010a,Liuetal2010b} and the
Harmonic Mean (HM) approximation developed by
\cite{Lugazetal2009,Lugazetal2010} (this combination is called the
HM triangulation for abbreviation in the following). The
triangulation results in the near Sun regions obtained by
\cite{Liuetal2013} will be adopted in this study as inputs of
prediction models. The corresponding prediction results are also
compared with the tracked kinematics of the CME front in the
subsequent Sun-to-Earth journey.

\subsection{2012 January 19 Event}
\subsubsection{Model Input}
The inputs of DGSTOA and DGSPM for this event are as follows. The
CME was launched at 13:55:00 UT on 2012 January 19, and it was
accelerated to max speed $V_M$=1362 km/s at the radial distance
$R_M$=15.67 $R_s$ at 18:45:54 UT ($t_M$) \citep{Liuetal2013}, which
is taken to be the start time of model predictions. The projected
angular width of this CME ($AW_{CME}$) is $360^{\circ}$ in the POS
of SOHO/LASCO. The solar wind speed ($V_{SW}$) at 1 AU is about 350
km/s observed by the WIND spacecraft at the CME launch time, and the
following ambient solar wind and IMF parameters at 1 AU are used to
compute $C_S$, $V_A$, and $V_F$: proton density $n$ = 5.8 cm$^{-3}$,
proton temperature $T_p = 0.7 \times 10^5$ K, IMF magnetude $B$ =
4.3 nT (see Figure 2). The parameters $V_M$, $R_M$, $t_M$, $V_{SW}$,
$C_S$, $V_A$, $V_F$ are input into Equations
(\ref{eq:vs_sstoa})--(\ref{eq:Ma}) for DGSTOA and Equations
(\ref{eq:vs_sspm})--(\ref{eq:essi}) for DGSPM. The angular width
$AW_{CME}$ is also required by DGSPM in order to predict whether or
not the shock will persist to the Earth.

In addition to DGSTOA and DGSPM, we also use the Drag-Based Model
(DBM) \citep{Vrsnaketal2013} to give predictions as a comparison.
This model assumes that beyond a certain distance the dynamics of
CMEs are governed solely by the interaction between interplanetary
CME (ICME) and the ambient solar wind, and the drag acceleration has
a quadratic dependence on their relative speed. This DBM model has
four free parameters, i.e. ``take-off speed'' $v_0$, starting radial
distance $r_0$, ambient solar wind speed $w$, and drag coefficient
$\gamma$. Here, we take $v_0$=$V_M$, $r_0$=$R_M$, $w$=$V_{SW}$. We
fit the distance-time plot predicted by the DBM with different
$\gamma$ to that of the HM triangulations from $R_M$ to $R_M$+50
$R_s$ (see Figure 1) to give the best drag coefficient, which is
$\gamma=9.8\times 10^{-8}$ km$^{-1}$ for this event. Then the DBM
with this finally determined value of $\gamma$ is adopted to predict
the propagation of the CME front beyond $R_M$+50 $R_s$.

\subsubsection{Observation and Prediction from the Sun to Earth}
Figure 1 shows a comparison between the observations of STEREO for
the CME front (HM triangulation) and predictions of the models
(DGSPM, DGSTOA, DBM) for the distance-time plot (top) and
velocity-distance plot (bottom). In this figure, the blue solid
circles are the tracked results of the HM triangulation for the CME
front. We can see that this CME underwent an initial impulsive
acceleration, and it reached the max speed 1362 km/s ($V_M$) at the
end of this acceleration phase at $R$ =15.67 $R_s$ ($R_M$); then,
the CME experienced a rapid deceleration from 15.67 $R_s$ to about
50 $R_s$; finally, it propagated outward with a gradual deceleration
\citep{Liuetal2013}. The application of the triangulation technique
requires that the CME front is observed by two STEREO spacecraft
simultaneously. For this case, the CME front was tracked by the
triangulation method continuously out to about 160 $R_s$, and a
linear extrapolation to 1 AU gave the predicted arrival time of the
shock at 03:11:00 UT on January 22 \citep{Liuetal2013}. Similarly,
the shock's propagation speed at the Earth predicted by the HM
triangulation was obtained by computing the average velocity during
the time when the CME front keeps nearly a constant speed. This
yielded 665 km/s for the shock speed \citep{Liuetal2013}. DGSPM and
DGSTOA are applicable beyond $R_M$. The DGSPM prediction is denoted
by the solid line, DGSTOA prediction denoted by the dashed line, and
DBM prediction denoted by the red dashed-dotted line. The
predictions of the two shock models (DGSPM, DGSTOA) are similar, and
differences between them are very small for this case. They both
roughly agree with the HM triangulation. Specifically speaking,
DGSPM predicts that the shock would have reached Earth at 21:14:04
UT on 2012 January 21, and the shock's propagation speed at the
Earth would be 653 km/s. DGSTOA predicts that the shock would have
encountered Earth at 21:53:12 UT on 2012 January 21 with the local
propagation speed of 623 km/s. According to DBM, the arrival time
and local speed at 1 AU would be 05:10:02 UT on 2012 January 23 and
383 km/s, respectively.

The in situ measurements of WIND at 1 AU demonstrated that the
corresponding shock arrived at Earth at 05:32:58 UT on January 22
\citep{Liuetal2013}, and the shock's local propagation speed along
the Sun-Earth direction was 466 km/s (see section 3.1.3) as shown by
green stars in Figure 1. Therefore, the prediction errors of DGSPM
for the shock's arrival time and local propagation speed at Earth
are 8.31 hours (hr) and -187 km/s, respectively. Similarly,
corresponding errors of DGSTOA are 7.66 hr and -157 km/s. As a
comparison, prediction errors of the HM triangulation are 2.37 hr
and -199 km/s. On the other hand, the kinematics of the CME front
beyond 40 $R_s$ predicted by DBM are evidently slower than the HM
triangulation. As a result, the DBM prediction fits the propagation
of the following ICME instead of the preceding shock. The arrival
time and local propagation speed of ICME at WIND are shown as green
triangles in Figure 1. The ICME speed (455 km/s) at 1 AU is defined
as the average speed across the ICME leading boundary (see section
3.1.3). The prediction errors of DBM for ICME arrival time and local
speed are -5.17 hr and 72 km/s, respectively.

In contrast with the HM triangulation, both shock models (DGSPM,
DGSTOA) did not reproduce the rapid deceleration phase of the CME
front between $R_M$ and 50 $R_s$. Two reasons may be responsible for
this deficiency. On one hand, the shock models adopted here are
single-fluid models, which ignore the energy loss due to
accelerating local particles at the shock front. The energy of the
energetic particles can be a large portion of the total kinetic
energy of the CME \citep{Mewaldtetal2008}. As envisioned by
\cite{Liuetal2013,Liuetal2016}, the energy loss to energetic
particles through shock acceleration may partly account for the
rapid deceleration. On the other hand, the solar wind speed is
assumed to be a constant from the Sun to Earth in these models,
which is taken from the in situ measurements of WIND at 1 AU. In
fact, the background solar wind should undergo a steady acceleration
through most of the distance within 30 $R_s$
\citep{Sheeleyetal1997}, and the corresponding solar wind speeds in
these regions are definitely slower than the 1 AU speed. The
overestimating of the ambient solar wind speed in these regions
increased the convection effect of the background flow, and weakened
the shock's rapid deceleration as a result. As to DBM, the input
parameter $\gamma$ was determined by fitting the predicted
distance-time plot to that of the HM triangulations in the range
from $R_M$ to $R_M$+50 $R_s$. Therefore, this model successfully
reproduced the rapid deceleration of the CME front during this
stage. However, DBM continued to predict a fast deceleration of the
CME front at large distances, and its prediction lagged both the
predictions of the shock models and the HM triangulation results for
the CME's leading edge more and more. As a result, the arrival time
predicted by DBM agreed better with that of the ICME, rather than
with that of the preceding shock. The propagation speed at 1 AU
predicted by DBM also agreed reasonably well with the average speed
across the ICME leading boundary.

\subsubsection{Comparison with In Situ Measurements}
Figure 2 displays the in situ observations of the WIND spacecraft
from 2012 January 19 to January 25. From top to bottom, the panels
in this figure display the proton density (n$_p$), bulk speed
(v$_p$), proton temperature (T$_p$), total magnetic field ($\mid$ B
$\mid$) and its three components (B$_x$, B$_y$, B$_z$) in the
Geocentric Solar Ecliptic (GSE) coordinate system. The vertical
dotted line denotes the CME launch time. The green dashed line
indicates the shock front. The arrival times predicted by DGSPM, and
DGSTOA are represented as the black solid line and black dashed
line, respectively. The blue dashed-dotted line is the arrival time
predicted by the HM triangulation. The shaded region is the ICME
structure following the shock \citep{Liuetal2013}. We define the
ICME local speed as the averaged flow speed across the ICME leading
boundary, which is 455 km/s for this event (see the pink line in the
bulk speed plot). The red dashed-dotted line represents the arrival
time predicted by DBM. This figure shows that the arrival times
predicted by DGSPM, DGSTOA, and HM triangulation are close to the
shock arrival time detected by WIND. But all predictions are earlier
than the real arrival time. The arrival time predicted by DBM is
late and falls inside the ICME interval, but close to the ICME
leading boundary.

The local propagation speed of the shock can be computed from the
upstream and downstream solar wind parameters:
\begin{eqnarray}\label{eq:R-H}
\textbf{V}_{sh}=\frac{\rho_2 \textbf{v}_2-\rho_1
\textbf{v}_1}{\rho_2-\rho_1}
\end{eqnarray}
Here $\rho_1$, $\rho_2$ are the average density in the upstream and
downstream regions; $\textbf{v}_1$, $\textbf{v}_2$ are the
corresponding average velocity; $\textbf{V}_{sh}$ is the velocity
vector of the shock front. As predictions of all these methods (HM
triangulation, DGSPM, DGSTOA) are made along the Sun-Earth line, we
compute the shock's propagation speed in the Sun-Earth direction,
i.e. $V_{sh}=\textbf{V}_{sh} \textbf{.} \textbf{n}_{SE}$. Here
$\textbf{n}_{SE}$ is the unit vector of the Sun-Earth direction.

As an example, Figure 3 demonstrates how we compute the shock's
local speed at 1 AU. We use the high-resolution (3 second) data of
the WIND observation to compute the upstream and downstream solar
wind parameters. Figure 3(a) displays solar wind parameters (n$_p$,
v$_p$, T$_p$, $\mid$ B $\mid$) in the upstream and downstream
regions of the shock at 05:32:58 UT on 2012 January 22. Here, time
is indicated relative to the shock front (green dashed line).
Negative time corresponds to the upstream region, and positive time
to the downstream region. The shock jump is very sharp in this case.
The solar wind parameters vary smoothly both upstream and
downstream. Figure 3(b) shows the computed shock speed $V_{sh}$ for
different averaging windows on either side of the shock front. We
find that the computed $V_{sh}$ increases gradually with growing
size of the averaging window from 0 to 3 minutes, then $V_{sh}$
decreases slightly from 3 to 6 minutes. After 6 minutes, $V_{sh}$
nearly stays constant at 466 km/s. There is no generally valid
standard to define the averaging window size. We take the saturation
level, here 466 km/s, as the in situ observed shock speed. The blue
solid lines in Figure 3(a) indicate the averages of solar wind
parameters $\rho_1$, $\rho_2$, $v_1$, $v_2$ in the upstream and
downstream regions with 6 minutes of averaging interval. The local
speed of the shock estimated above is slightly higher than the
averaged flow speed (415 km/s) in the sheath region between the
shock and ICME, but is closer to the predicted values of models (665
km/s of HM triangulation, 653 km/s of DGSPM, 623 km/s of DGSTOA).
For clarity, Table 1 lists the input \& output parameters of the
models (HM, DGSPM, DGSTOA, DBM) and the in situ observations for
this case.

For a brief summary to this case, predictions of the two shock
models (DGSPM, DGSTOA) did not reproduce the rapid deceleration
phase of the CME front. But they ``kept up'' with the shock in the
following journey, and predicted both the arrival time and
propagation speed of the shock at Earth reasonably well. Although
DBM successfully reproduced the rapid deceleration of the CME, it
lagged the shock front more and more in the subsequent propagation.
As a result, the prediction of DBM pointed to the ICME behind the
shock. The inputs of DGSPM and DGSTOA are the kinematical parameters
of the CME front when it was accelerated to maximum speed, while the
free input parameter $\gamma$ of DBM needs to be determined by
fitting to the propagation process of the CME front during its
deceleration phase. Therefore, the prediction results of DGSPM and
DGSTOA were obtained earlier than those of DBM. All inputs of these
models were obtained from the HM triangulation results based on
STEREO observations.

\subsection{2012 January 23 Event}
\subsubsection{Model Input}
The inputs of DGSTOA and DGSPM for this event are as follows. The
CME was launched at 03:40 UT on 2012 January 23, and it was
accelerated to max speed $V_M$=1542 km/s at the radial distance
$R_M$=12.24 $R_s$ at 05:35:54 UT ($t_M$) \citep{Liuetal2013}, which
is the start time of the model predictions. $AW_{CME}$ is
$360^{\circ}$ in the POS of SOHO/LASCO. $V_{SW}$ at 1 AU is about
460 km/s detected by the WIND spacecraft at the CME launch time, and
the following ambient solar wind and IMF parameters at 1 AU are used
to compute $C_S$, $V_A$, and $V_F$: $n$ = 10.0 cm$^{-3}$, $T_p = 0.3
\times 10^5$ K, $B$ = 11.1 nT (see Figure 5). The parameters $V_M$,
$R_M$, $t_M$, $V_{SW}$, $C_S$, $V_A$, $V_F$ are input into Equations
(\ref{eq:vs_sstoa})--(\ref{eq:Ma}) for DGSTOA and Equations
(\ref{eq:vs_sspm})--(\ref{eq:essi}) for DGSPM. $AW_{CME}$ is also
adopted by DGSPM to predict whether or not the shock will persist to
the Earth. We take $v_0$=$V_M$, $r_0$=$R_M$, $w$=$V_{SW}$ for the
DBM model, and derive the drag coefficient $\gamma=3.4\times
10^{-8}$ km$^{-1}$ by fitting the model predictions to STEREO
observations for the CME front (HM triangulation results) in the
distance range from $R_M$ to $R_M$+50 $R_s$. Then, the DBM with this
finally determined value of $\gamma$ is adopted to predict the
propagation of the CME beyond $R_M$+50 $R_s$.

\subsubsection{Observation and Prediction from the Sun to Earth}
Figure 4 demonstrates a comparison between the observation of STEREO
and the predictions of the models (DGSPM, DGSTOA, DBM) for the
distance-time plot (top) and velocity-distance plot (bottom) of the
CME front. Definitions of lines and symbols in this figure are the
same as those in Figure 1. After the impulsive acceleration, this
CME underwent a clear deceleration phase from $R_M$ to about 70
$R_s$. But this deceleration is not as strong as for the 2012
January 19 CME event shown in Figure 1. For example, the propagation
speed of the CME decreased from 1542 km/s only to 900 km/s when
distance increased from $R_M$ to 90 $R_s$. The shock models (DGSPM,
DGSTOA) match reasonably well the HM triangulation results during
this deceleration stage. Beyond 70 $R_s$, the HM triangulation
yields a nearly constant-speed propagation of the CME front, and the
shock models predict a gradual deceleration. The prediction of DGSPM
is more consistent with the HM triangulation than DGSTOA during this
phase. DBM predicts a weaker deceleration within 70 $R_s$ than DGSPM
and DGSTOA, but a stronger deceleration beyond 70 $R_s$. For this
case, the HM triangulation tracked the CME front continuously out
from the Sun to beyond the Earth orbit, and it predicted that the
shock would have reached Earth at 23:06:00 UT on 2012 January 24
with the local propagation speed of 900 km/s at 1 AU
\citep{Liuetal2013}. DGSPM predicts that the shock would have
arrived at Earth at 01:13:15 UT on 2012 January 25 with the local
propagation speed of 799 km/s. DGSTOA predicts that the shock would
have encountered Earth at 03:40:06 UT on 2012 January 25 with the
local speed of 718 km/s. DBM predicts that the ICME would have
reached Earth at 06:59:48 UT on 2012 January 25 with the local speed
of 604 km/s at 1 AU.

The in situ measurements of WIND at 1 AU demonstrated that the
corresponding shock arrived at Earth at 14:40:06 UT on January 24
\citep{Liuetal2013}, and the shock's local propagation speed along
the Sun-Earth direction was 719 km/s (see section 3.2.3), shown as
green stars in Figure 4. The shock arrived earlier than all
predictions. The best prediction for the arrival time comes from the
HM triangulation with the error of -8.43 hr. The prediction errors
of the shock arrival time are -10.55 hr and -13.0 hr for DGSPM and
DGSTOA, respectively. The fast propagation of the shock for this
case might be due to its interaction with an earlier event as
pointed out by \cite{Liuetal2013}. Although models predict delayed
arrival times, the shock's local propagation speeds at Earth
predicted by them are very close to the speed computed from the in
situ observations. The predictions errors are -181 km/s for HM, -80
km/s for DGSPM, and 1 km/s for DGSTOA. No ICME structure was found
in the WIND observations for this shock event \citep{Liuetal2013}.
But we can see a clear sheath region downstream the shock with high
fluctuations in solar wind parameters. The sheath structure ended at
about 12:00 UT of January 25, and the average speed across its back
boundary was around 602 km/s. If we relate the prediction of DBM to
the sheath's back boundary (SBB), then the prediction errors for
arrival time and 1 AU local speed are only 5.0 hr and -2 km/s.

\subsubsection{Comparison with In Situ Measurements}
Figure 5 displays the in situ observations of the WIND spacecraft
from 2012 January 23 to January 29. Definition of each line in this
figure is the same as in Figure 2. The green dashed-dotted line
denotes the back boundary of the sheath structure. The pink line in
the bulk speed plot represents the averaged flow speed across the
sheath's tail part. The shock arrival time (green dashed line) is
earlier than prediction of any model. We need to point out that this
shock had propagated into the ICME structure associated with the
January 19 CME event, and the interaction should have occurred
inside 1 AU. This shock formed the back boundary of the preceding
ICME. The former ICME offered a low density and strong magnetic
field background medium for the propagation of this shock. Both low
density and strong magnetic field increased the Alfv\'en speed of
the ambient medium, which would cause a faster advance of the shock
than predicted. A similar analysis can be found in
\cite{Liuetal2013}. The arrival time of the ICME predicted by DBM is
5 hours earlier than the ending time of the sheath structure. In the
same way as for the 2012 January 19 event, we compute the in situ
speed of the shock along the Sun-Earth line based on the WIND
observations, which is 719 km/s. Similarly, this local speed of the
shock is slightly higher than the averaged flow speed (650 km/s) in
the sheath region, and is closer to the predicted values of models
(900 km/s of HM triangulation, 799 km/s of DGSPM, 718 km/s of
DGSTOA) as well. The detailed input \& output parameters of the
models (HM, DGSPM, DGSTOA, DBM) as well as the in situ observations
for this case are also listed in Table 1.

As a brief summary, this CME decelerated more slowly during the
rapid deceleration phase than the 2012 January 19 CME. Shock models
(DGSPM, DGSTOA) successfully reproduced this deceleration process
for this case. However, the shock propagated faster at larger
distances than all model predictions possibly due to the interaction
with the earlier ICME. As a result, the actual shock arrival time
was earlier than those predicted by DGSPM and DGSTOA. But the
shock's local propagation speed at 1 AU computed from the upstream
and downstream solar wind parameters matched predictions of the
models reasonably well, and this local speed of the shock was faster
than the sheath region flow speed. The corresponding ICME associated
with this shock was not recorded by WIND at 1 AU. The prediction of
DBM pointed to the back boundary of the sheath structure.

\subsection{2012 March 7 Event}
\subsubsection{Model Input}
This CME was launched at 00:15 UT on 2012 March 7, and it was
accelerated to max speed $V_M$=2369 km/s at the radial distance
$R_M$=15.20 $R_s$ at 02:36:36 UT ($t_M$) \citep{Liuetal2013}. This
time is taken to be the start time of shock models. $AW_{CME}$ is
$360^{\circ}$ in the POS of SOHO/LASCO, $V_{SW}$ at 1 AU is about
375 km/s detected by the WIND spacecraft at the CME launch time, and
the following ambient solar wind and IMF parameters at 1 AU are used
to compute $C_S$, $V_A$, and $V_F$: $n$ = 5.8 cm$^{-3}$, $T_p = 0.3
\times 10^5$ K, $B$ = 8.3 nT (see Figure 7). The parameters $V_M$,
$R_M$, $t_M$, $V_{SW}$, $V_A$, $C_S$, $V_F$ are input into Equations
(\ref{eq:vs_sstoa})--(\ref{eq:Ma}) for DGSTOA and Equations
(\ref{eq:vs_sspm})--(\ref{eq:essi}) for DGSPM. $AW_{CME}$ is also
required by DGSPM in order to predict whether or not the shock will
persist to the Earth. We take $v_0$=$V_M$, $r_0$=$R_M$, $w$=$V_{SW}$
for DBM, and derive the drag coefficient $\gamma=6.2\times 10^{-8}$
km$^{-1}$ by fitting the model predictions to STEREO observations
for the CME front (HM triangulations) in the distance range from
$R_M$ to $R_M$+50 $R_s$. Finally, the DBM with this finally
determined value of $\gamma$ is used to predict the propagation of
the CME beyond $R_M$+50 $R_s$.

\subsubsection{Observation and Prediction from the Sun to Earth}
Figure 6 shows observations of STEREO and predictions of the models
(DGSPM, DGSTOA, DBM) for the distance-time plot (top) and
velocity-distance plot (bottom) of the CME front. Definitions of
lines and symbols in this figure are the same as those in Figure 1.
Similar to the 2012 January 19 CME event, this CME underwent a rapid
deceleration from $R_M$ to about 50 $R_s$. The deceleration occurred
over a distance of only 35 $R_s$, but the velocity decreased from
2369 km/s to about 1000 km/s. Again, the shock models (DGSPM,
DGSTOA) could not reproduce this rapid deceleration, but the shock
velocity evolution predicted by them are consistent with the HM
triangulation results during the gradual deceleration phase beyond
100 $R_s$. The DBM successfully reproduced the rapid deceleration
from $R_M$ to 50 $R_s$. But it gave a too slow propagation beyond 60
$R_s$. For this case, the CME front was tracked by the HM
triangulation continuously out only to 150 $R_s$; the arrival time
of the shock and its propagation speed at 1 AU predicted by HM could
be determined by linear extrapolations, which gave 17:55:00 UT on
2012 March 8 and 950 km/s, respectively \citep{Liuetal2013}. DGSPM
predicts that the shock would have arrived at Earth at 13:34:18 UT
on 2012 March 8 with the local propagation speed of 873 km/s. DGSTOA
predicts that the shock would encountered Earth at 12:12:36 UT on
2012 March 8 with the local speed of 905 km/s. DBM predicts that the
ICME would have reached Earth at 17:24:36 UT on 2012 March 9 with
the local speed of 444 km/s.

The in situ measurements of WIND at 1 AU demonstrated that the
corresponding shock encountered Earth at 10:30:45 UT on March 8
\citep{Liuetal2013}, and the local propagation speed of the shock
along the Sun-Earth direction was 1088 km/s (see section 3.3.3),
shown as green stars in Figure 6. The arrival times predicted by the
shock models (DGSPM, DGSTOA) are very close to the observed one, and
their prediction errors are only -3.06 hr for DGSPM and -1.70 hr for
DGSTOA. The corresponding errors in the local shock speed are 215
km/s (DGSPM) and 183 km/s (DGSTOA), respectively. Predictions of the
HM triangulation are worse for the arrival time with an error of
-7.40 hr but better for the shock speed with an error of 138 km/s.
The associated ICME arrived at 1 AU at 05:00 UT on March 9 and
lasted until 12:20 UT on March 11 \citep{Liuetal2013}. The averaged
flow speed across the ICME leading boundary was 717 km/s (see
section 3.3.3). The predictions of DBM are 12.41 hr late for the
arrival time and fall short by 273 km/s for the ICME speed at 1 AU.

\subsubsection{Comparison with In Situ Measurements}
Figure 7 shows the in situ observations of the WIND spacecraft from
2012 March 6 to March 12. Definition of each line in this figure is
the same as that in Figure 2. The shock arrival time (green dashed
line) is earlier than the predictions of the models (HM, DGSPM,
DGSTOA). But the differences between them are small. The shaded
region is the interval of the following ICME structure
\citep{Liuetal2013}, which has a low density and temperature,
decreasing velocity, enhanced magnetic field strength, and a smooth
rotation of the magnetic field direction. The arrival time predicted
by DBM lies within the ICME interval, 12.41 hr later than the
arrival of the ICME leading boundary. In the same way, we compute
the in situ speed (1088 km/s) of the shock along the Sun-Earth
direction based on the WIND observations. Similarly, this shock
speed is faster than the averaged flow speed (680 km/s) observed in
the sheath region, and is closer to the predicted values of models
(950 km/s of HM triangulation, 873 km/s of DGSPM, 905 km/s of
DGSTOA). Table 1 summarizes the input \& output parameters of the
models (HM, DGSPM, DGSTOA, DBM) and the in situ observations for
this case.

For a summary, the rapid deceleration of this CME front from $R_M$
to 50 $R_s$ was successfully reproduced by DBM, but not by the shock
models (DGSPM, DGSTOA). The velocity evolutions predicted by DGSPM
and DGSTOA matched the HM triangulation results in the subsequent
gradual deceleration phase, and they gave reasonably good
predictions not only for the shock arrival time but also for the
shock's local speed at Earth. The predictions of DBM corresponded to
the ICME structure following the shock, but the predicted arrival
time lagged the ICME leading boundary by 12.41 hr. The shock speed
computed from the upstream and downstream solar wind parameters
described better the local propagation speed of the shock at 1 AU
than the flow speed in the sheath region. The speeds predicted by
the shock models were more consistent with the shock speed thus
derived.

\section{SUMMARY AND DISCUSSION}
In this study we employed two models to predict the propagation of
shocks associated with fast CMEs in the heliosphere, i.e. DGSTOA and
DGSPM. The former is based on the similarity theory of blast waves
in the ambient flow (solar wind) reference frame, and the latter on
the non-similarity theory of blast waves in the stationary reference
frame. The inputs of these models are kinematical characteristics of
the associated CME front when the CME reaches its maximum speed
close to the Sun, and these parameters are obtained from a
continuous tracking for the white-light feature of a CME viewed in
coordinated imaging of STEREO by the HM triangulation
\citep{Liuetal2013}. We applied these models to three cases of the
2012 January 19 (Case 1), January 23 (Case 2), and March 7 (Case 3)
CMEs/shocks, and found that the models can give a reasonable
prediction for the propagation of the shock associated with fast
CMEs after the initial impulsive acceleration phase. The shock's
arrival time and local speed at Earth predicted by the models match
generally well the in situ observations of WIND. We also used DBM to
predict the arrival time and local speed at Earth, and found that
its predictions agree better with the following ICME (Case 1 and
Case 3) or sheath structure (Case 2) rather than with the preceding
shock.

Table 2 summarizes the mean of the absolute prediction errors for
the CME/shock arrival times and their local propagation speeds at
Earth predicted by the HM triangulation, DGSPM, DGSTOA, and DBM. We
can see that the mean of prediction errors for arrival times ranges
from 6.0 to 7.6 hr, and the mean error for propagation speeds ranges
from 110 to 180 km/s. The prediction accuracies of different models
are similar. These predictions are reasonably good as far as we
know. For example, \cite{zhaoanddryer2014} presented a review of
current status of the prediction models for the CME/shock arrival
time, and found that their prediction errors are about 10-12 hr at
present. The prediction accuracies of arrival time derived in this
study are improved by about 4-5 hr compared with them. Prediction
results of DGSPM and DGSTOA are obtained earlier than those of DBM
due to no free input parameter in the formers. The HM triangulation
technique can provide a description about the whole propagation
process of the CME front, including the impulsive acceleration,
rapid deceleration, and slow deceleration phases
\citep{Liuetal2013}. However, we need to point out that the shock
models (DGSPM, DGSTOA) provided by this study may not apply to slow
CMEs with initial speeds lower than the ambient solar wind speed.
This kind of CMEs usually undergoes a gradual acceleration in the
beginning, and then propagates outward at nearly a constant speed
\citep{Liuetal2016}.

One reason for the prediction errors of the shock models may come
from the uncertainties in the input parameters. For example, the
maximum speed $V_M$ of the CME front is one of the most important
input parameters. This parameter has uncertainties shown as blue
error bars in Figure 1, Figure 4, and Figure 6. Specific speaking,
the relative uncertainties of $V_M$ are 22.4\% (305/1362) for Case
1, 11.8\% (182/1542) for Case 2, and 14.0\% (331/2369) for Case 3,
respectively. As a comparison, the relative errors of the shock
arrival time predicted by DGSPM are 13.1\% (8.31/63.63), 30.1\%
(10.55/35.0), 8.9\% (3.06/34.26), and those of the local shock speed
at 1 AU are 40.1\% (187/466), 11.1\% (80/719), 19.8\% (215/1088).
The solar wind speed, adopted from 1 AU observations at the CME
launch time, also does not represent the real flow speed just
upstream of the shock. Considering so many input uncertainties, the
prediction errors are acceptable. The shock speed computed from the
upstream and downstream solar wind parameters is more consistent
with the values predicted by the shock models than the average speed
in the sheath.

The propagation processes of all three cases consist of three phases
from the Sun to Earth: an initial impulsive acceleration, a rapid
deceleration, and finally a gradual deceleration
\citep{Liuetal2013}. The impulsive acceleration is caused by the
Lorentz force, and ends up at a distance of 10-20 $R_s$. Predictions
of DGSPM and DGSTOA start when the CME front reaches its maximum
speed. But prediction of DBM starts at 50 $R_s$ (set empirically)
further out because the drag coefficient ($\gamma$) needs to be
determined by fitting. The rapid deceleration occurs over a very
short distance from 10-20 $R_s$ to about 50 $R_s$. The shock models
adopted here can not reproduce this rapid deceleration (for Case 1
and Case 3, Case 2 is better). An overestimation of the solar wind
speed (taken from 1 AU), and therefore a too large convection of the
ambient flow adopted in these models, can be one potential reason.
Besides this, the energy loss due to accelerating the local
energetic particles by the shock can be another reason
\citep{Liuetal2013}, as this energy loss is not taken into account
of in these single-fluid shock models. For example,
\cite{Mewaldtetal2008} estimated that the total content of energy
for energetic particles produced by shock can account for 10\% of
the associated CME's kinetic energy, or even more. As a contrast,
DBM successfully reproduces this rapid deceleration as we fit the
DBM predictions to the HM triangulation results in these regions to
determine its free drag coefficient. The rapid deceleration evolves
to a gradual deceleration at a distance around 50 $R_s$. Predictions
of DGSPM and DGSTOA agree with the kinematics of the CME front
tracked by the HM triangulation in this phase. The predicted shock
arrival time and local propagation speed at Earth are consistent
with the in situ measurements of WIND. However, the DBM predicts a
stronger deceleration than the HM triangulation during this gradual
deceleration stage, which indicates that a different $\gamma$ value
may be needed for the gradual deceleration. This result implies that
different physical mechanisms are responsible for the different
deceleration processes, which cannot be described by a single
drag-based deceleration process. The DBM predicted result lags the
shock propagation more and more (see Figure 1, Figure 4, and Figure
6). At 1 AU, it's prediction agrees with the leading boundary of the
ICME (Case 1 and Case 3) or the back boundary of the sheath (Case 2)
behind the shock for both the arrival time and the local propagation
speed. This may indicate that the standoff distance between the ICME
and its preceding shock is increasing during their outward
propagation.

It will lead to another point to be clarified: what is the nature of
the CME front in the wide-angle imaging observations? As pointed out
by \cite{Liuetal2011,Liuetal2013}, the density within the CME is
stronger than that of the preceding shock and sheath close to the
Sun, therefore the white-light feature in imaging observations in
these regions represents the front boundary of the CME main body;
far away from the Sun, the density compression of the shock
eventually dominates over that at the CME main body due to fast
expansion of the CME, and the white-light feature gradually shifts
to the preceding shock structure as a result. More direct evidences
on this point will be investigated in future studies.

\acknowledgments This work is jointly supported by the National
Basic Research Program (973 program) under grant 2012CB825601 and
2012CB957801, the National Natural Science Foundation of China
(41231068, 41274179, 41474153, 41374173, and 41274192), the
Specialized Research Fund for State Key Laboratories of China, the
visiting program of China Scholarship Council with CSC No.
201404910005, and the Youth Innovation Promotion Association of
Chinese Academy of Sciences under Grant No. 2016133. X. H. Zhao is
also supported by the MPS fund. Y. D. Liu is also supported by the
Recruitment Program of Global Experts of China. We acknowledge the
use of data from STEREO, WIND, and SOHO.

\clearpage

\begin{figure}
\epsscale{1.0} \plotone{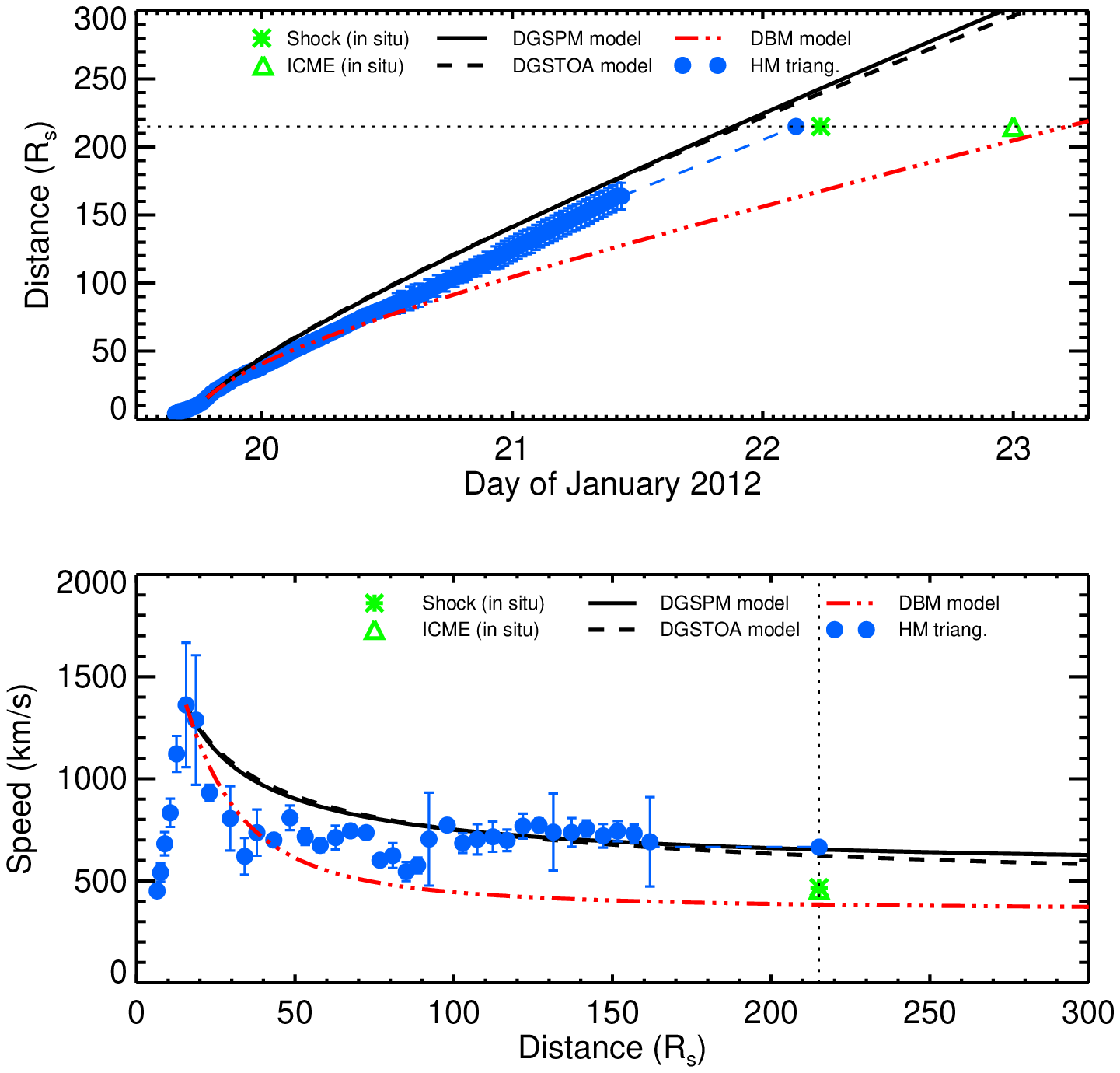} \caption{Comparison between
imaging observations of STEREO and model predictions for the
distance-time plot (up) and velocity-distance plot (bottom) for the
2012 January 19 event. The blue solid circles with error bars
represent the HM triangulation results based on imaging observations
of STEREO \citep{Liuetal2013}. The black solid and dashed lines
denote the predictions of DGSPM and DGSTOA, respectively. The red
dashed-dotted line denotes the prediction of DBM. The green star and
triangle stand for the in situ measurements of WIND at 1 AU for the
shock and ICME. The dotted lines denote the Earth
location.\label{fig1}}
\end{figure}
\clearpage

\begin{figure}
\epsscale{0.85} \plotone{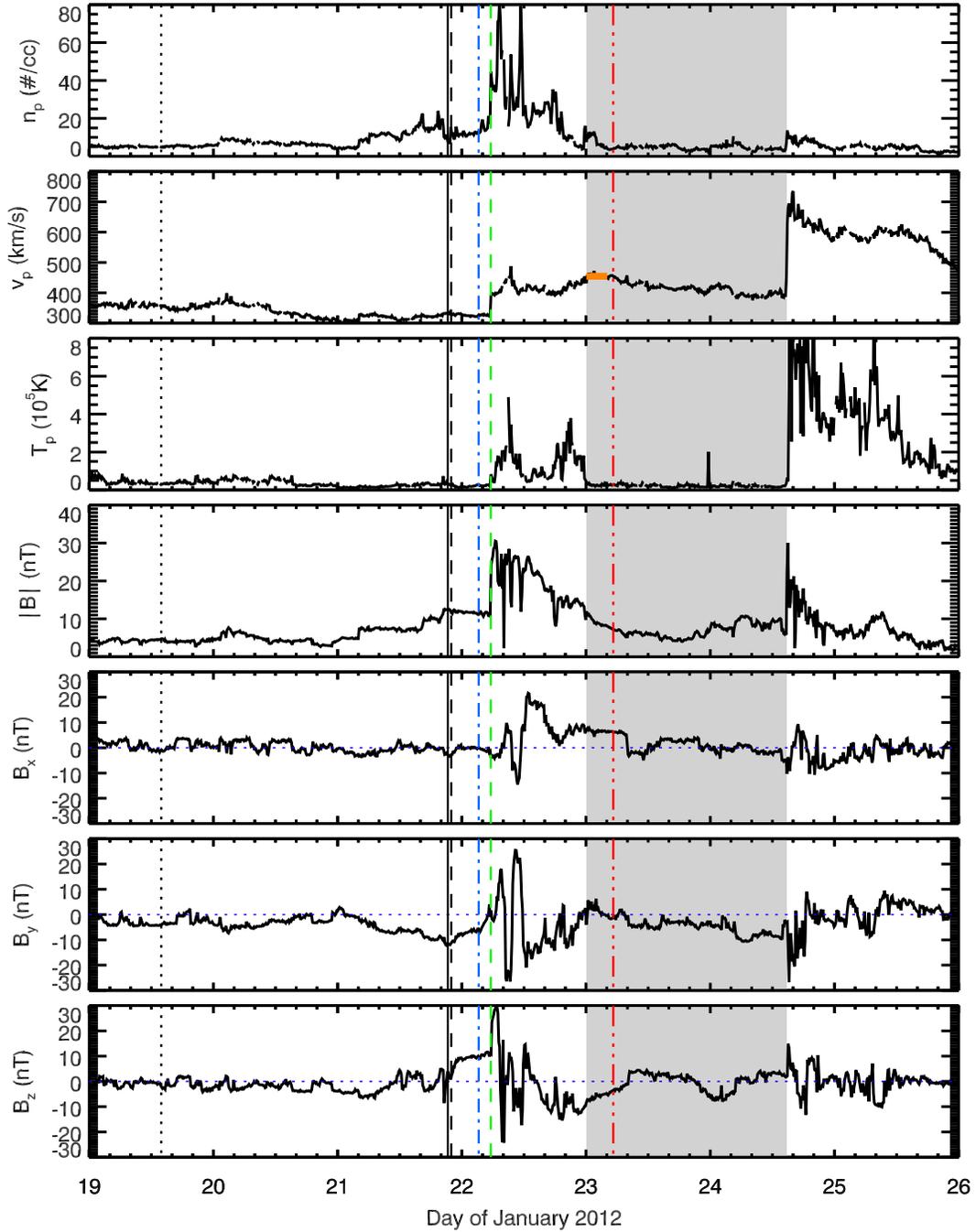} \caption{The in situ observations
of the WIND spacecraft from 2012 January 19 to 2012 January 25. The
panels display the proton density (n$_p$), bulk speed (v$_p$),
proton temperature (T$_p$), total magnetic field ($\mid$ B $\mid$)
and its three components (B$_x$, B$_y$, B$_z$) in the GSE coordinate
system from top to bottom. The vertical dotted line denotes the CME
launch time, the vertical green dashed line denotes the shock, and
the shaded region denotes the ICME interval. The times of the shock
arrival and ICME interval are taken from \cite{Liuetal2013}. The
arrival times predicted by DGSPM, DGSTOA, HM, and DBM are indicated
by the vertical solid line, dashed line, blue dashed-dotted line,
and red dashed-dotted line, respectively. The horizontal pink line
in the bulk speed plot represents the average speed across the ICME
leading boundary.\label{fig2}}
\end{figure}
\clearpage

\begin{figure}
\epsscale{0.85} \plotone{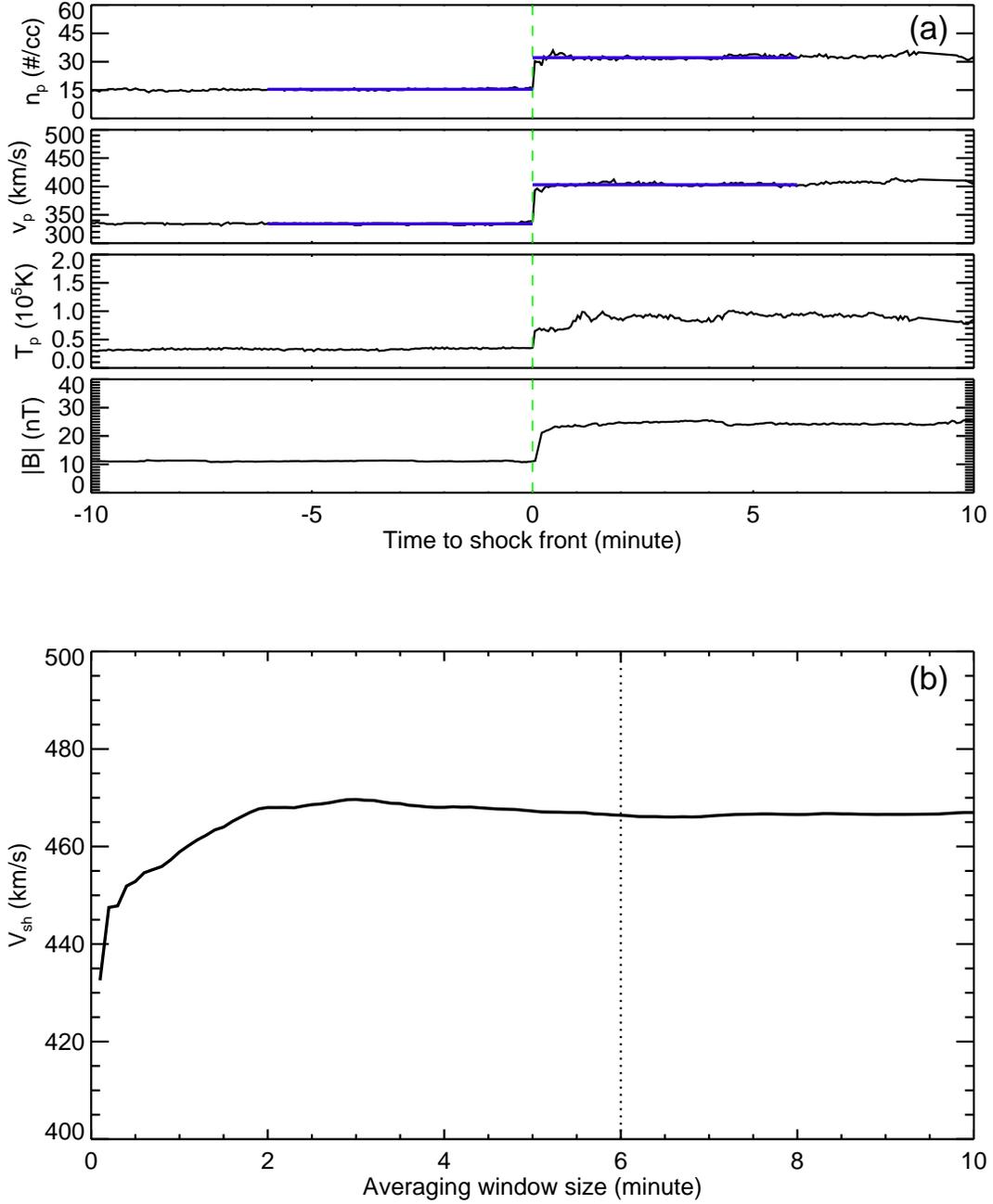} \caption{The upstream and
downstream solar wind parameters (n$_p$, v$_p$, T$_p$, $\mid$ B
$\mid$) of 3-second resolution for the shock at 05:32:58 UT on 2012
January 22 (a) and the shock propagation speed ($V_{sh}$) plotted
against the averaging window size in the upstream and downstream
regions. In (a), the blue solid lines indicate the averages of solar
wind density and velocity in the upstream and downstream regions. In
(b), the vertical dotted line corresponds to 6 minutes of the
averaging window size finally selected.\label{fig3}}
\end{figure}
\clearpage

\begin{figure}
\epsscale{1.0} \plotone{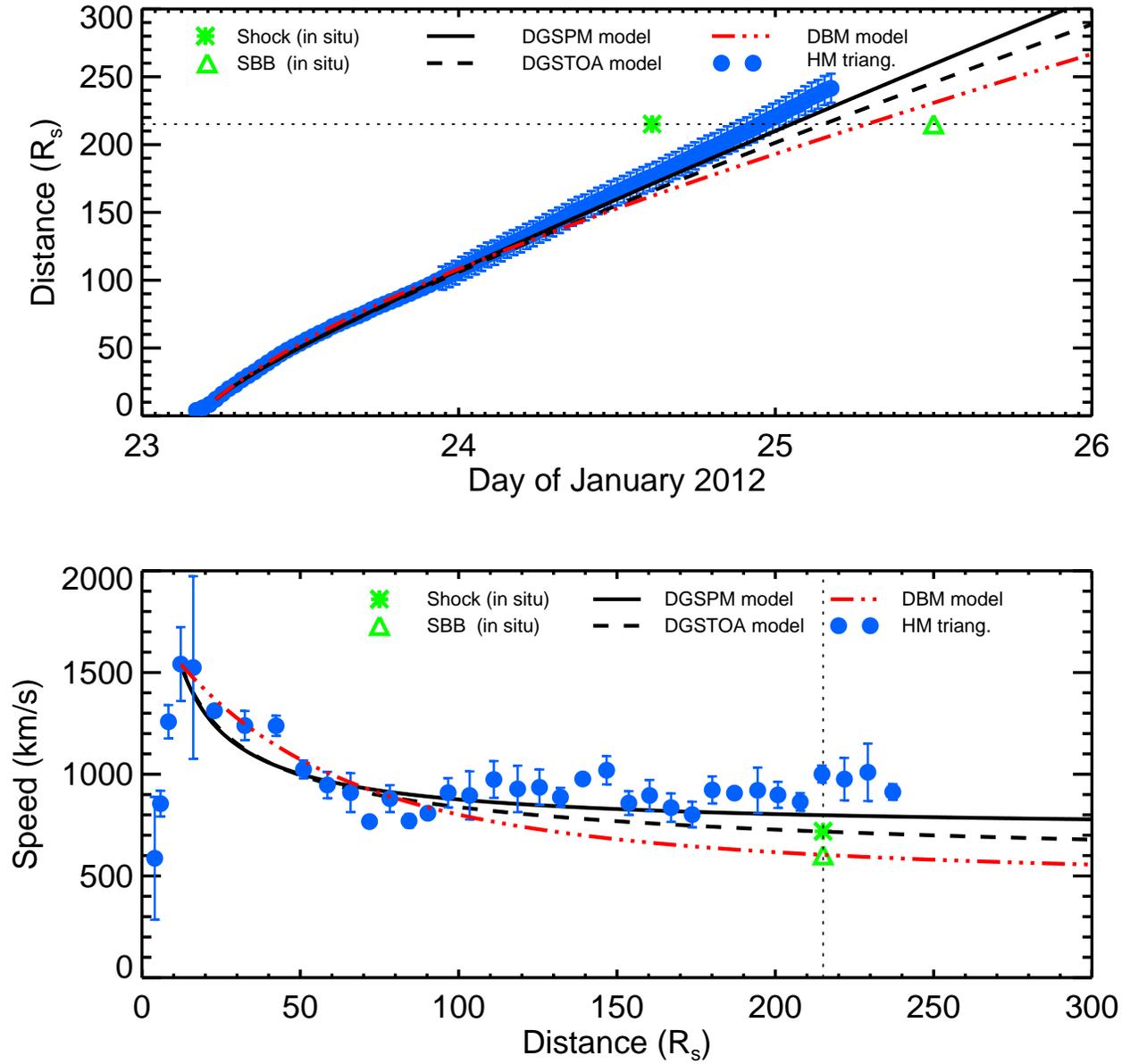} \caption{Similar to Figure 1, but
for the 2012 January 23 CME event. SBB stands for the sheath's back
boundary.\label{fig4}}
\end{figure}
\clearpage

\begin{figure}
\epsscale{0.85} \plotone{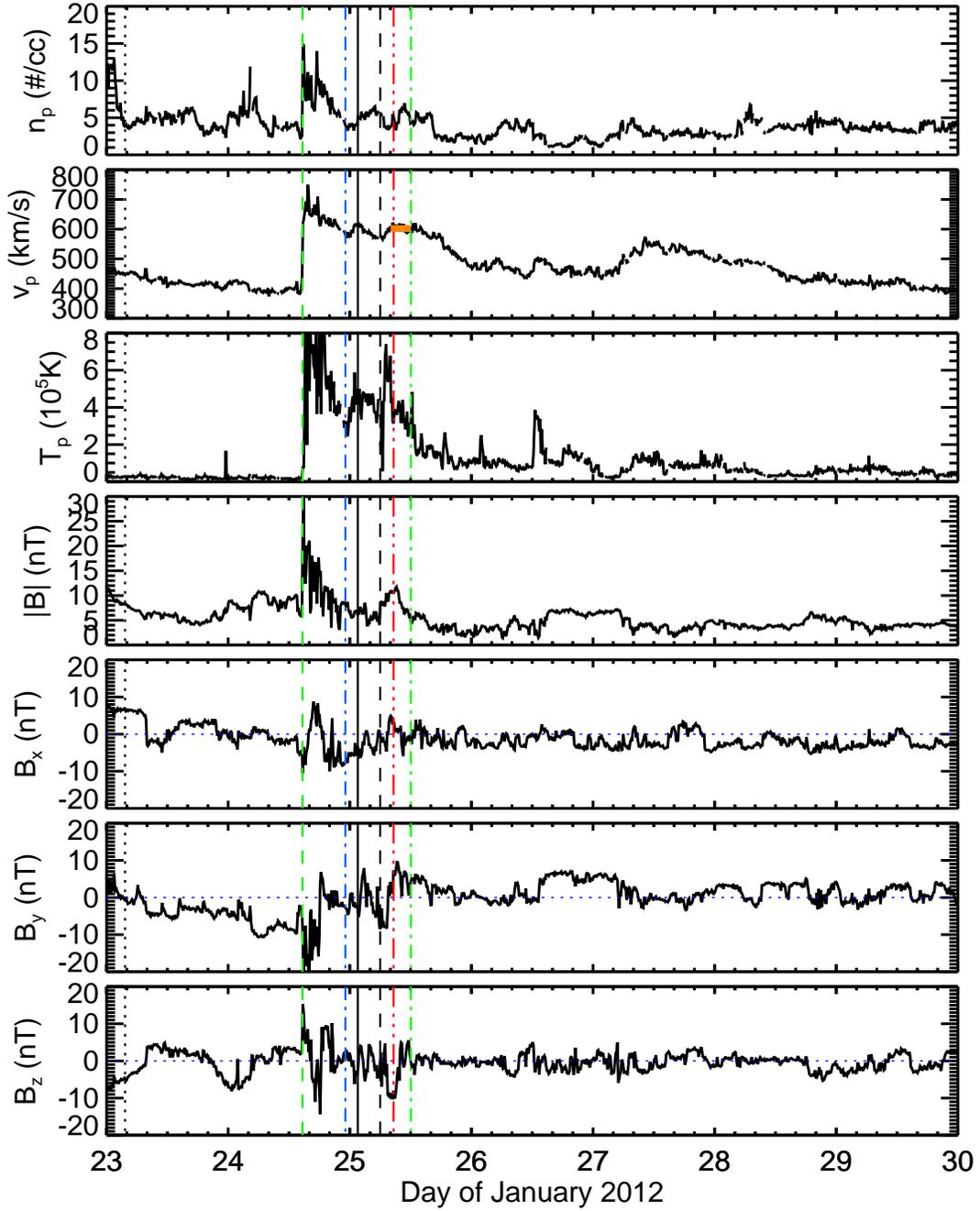} \caption{Similar to Figure 2, but
for the in situ observations of WIND from 2012 January 23 to January
29, including the 1 AU observations for the 2012 January 23 CME
event. No ICME signatures are observed at WIND except the shock
\citep{Liuetal2013}, and the shock arrival time is taken from
\cite{Liuetal2013}. The green dashed-dotted line denotes the back
boundary of the sheath structure. The pink line in the bulk speed
plot represents the average flow speed across the sheath's tail
part.\label{fig5}}
\end{figure}
\clearpage

%\begin{figure}
%\epsscale{1.0} \plotone{fig6.eps} \caption{Similar as Figure 3, but
%for the shock at 14:40:06 UT on 2012 January 24.\label{fig6}}
%\end{figure}
%\clearpage

\begin{figure}
\epsscale{1.0} \plotone{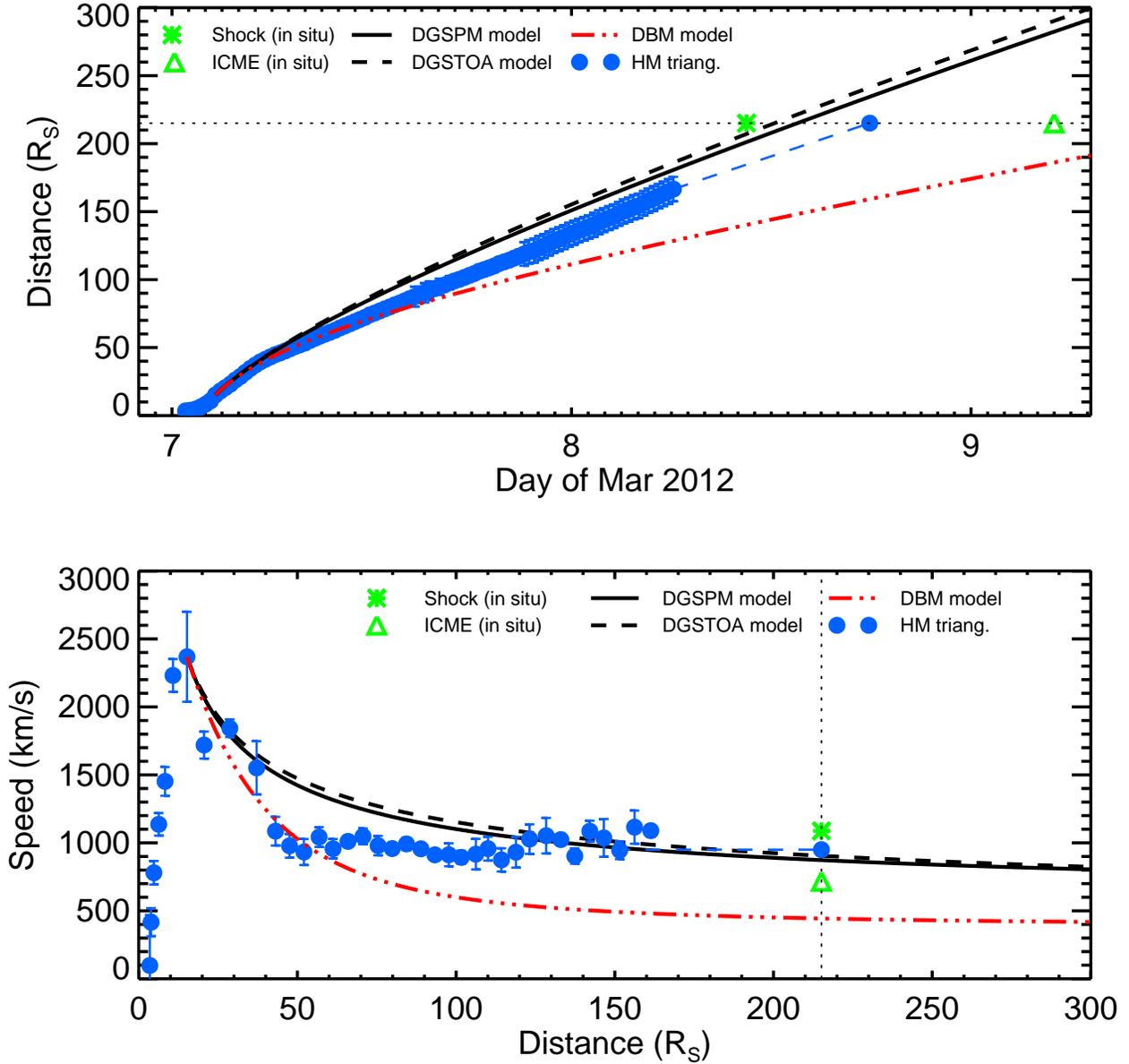} \caption{Similar to Figure 1, but
for the 2012 March 7 CME event.\label{fig6}}
\end{figure}
\clearpage

\begin{figure}
\epsscale{0.85} \plotone{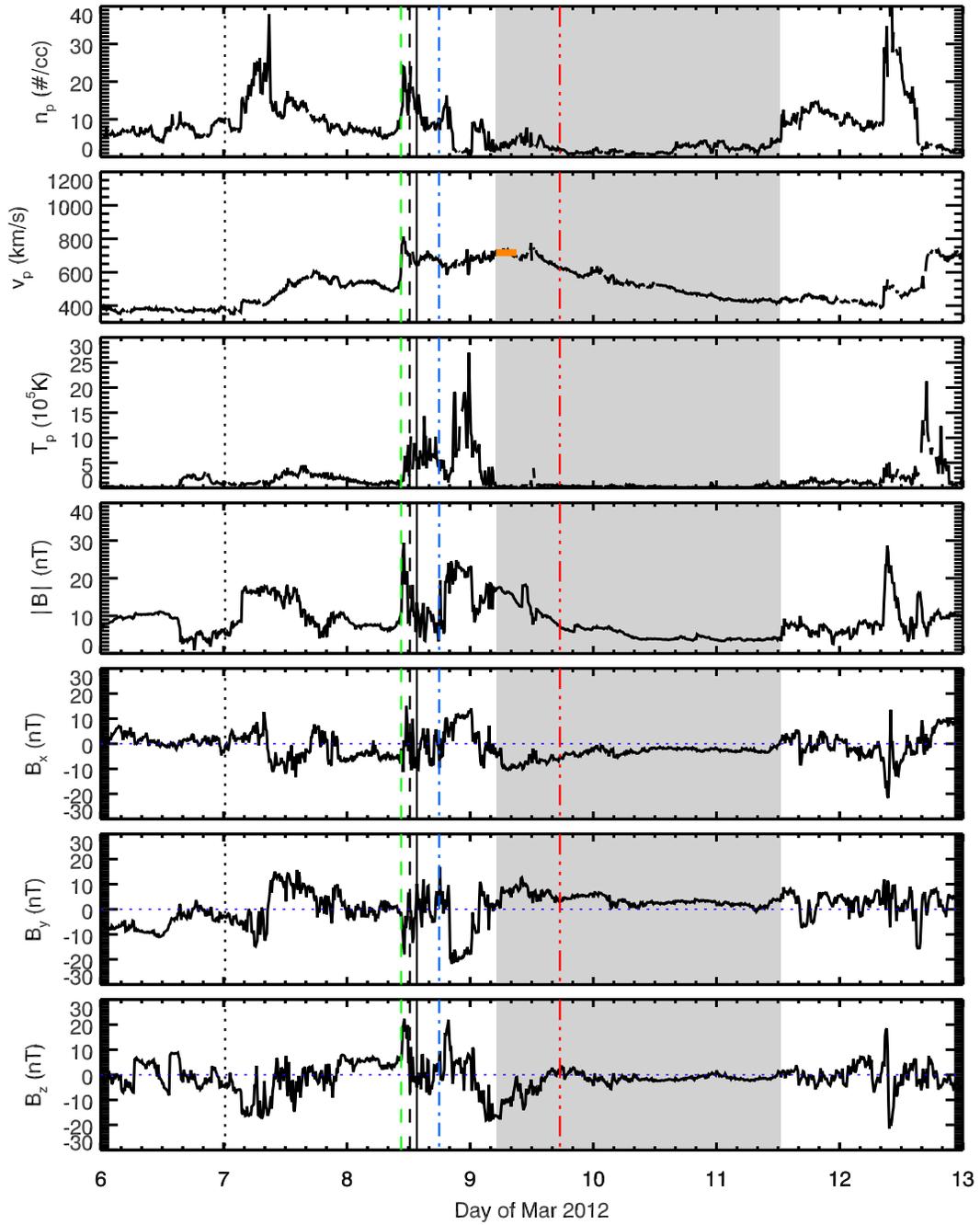} \caption{Similar to Figure 2, but
for the in situ observations of WIND from 2012 March 6 to March 12,
including the 1 AU observations for the 2012 March 7 CME event. The
times of the shock arrival and ICME interval are taken from
\cite{Liuetal2013}. The pink line in the bulk speed plot represents
the average speed across the ICME leading boundary.\label{fig7}}
\end{figure}
\clearpage

%\begin{figure}
%\epsscale{1.0} \plotone{fig9.eps} \caption{Similar as Figure 3, but
%for the shock at 10:30:45 UT on 2012 March 8.\label{fig9}}
%\end{figure}
%\clearpage

\begin{table}\scriptsize %\tiny\scriptsize\footnotesize\small\normalsize
\caption{The inputs \& outputs of models, and in situ observations
for these three cases.} \centering
\begin{tabular}{c|c|c|c|c}
\hline \hline
 \multicolumn{2}{c|}{}  & Case 1 & Case 2 & Case 3 \\
\hline
\multicolumn{2}{c|}{CME launch time} & 2012.1.19 13:55:00 UT & 2012.1.23 03:40:00 UT&2012.3.7 00:15:00 UT \\
\hline
\multirow{6}{*}{Inputs}  & $t_M$& 2012.1.19 18:45:54 UT &2012.1.23 05:35:54 UT & 2012.3.7 02:36:36 UT \\
  &  $R_M$ & 15.67 $R_s$  & 12.24 $R_s$ & 15.20 $R_s$\\
  &  $V_M$ & 1362 km/s & 1542 km/s & 2369 km/s \\
  &  $AW_{CME}$  & $360^{\circ}$ & $360^{\circ}$ & $360^{\circ}$ \\
  & $V_{SW}$ & 350 km/s & 460 km/s & 375 km/s \\
  & $V_{A}$ & 39 km/s & 77 km/s & 75 km/s \\
  & $C_{S}$ & 35 km/s & 23 km/s & 23 km/s \\
  & $V_{F}$ & 49 km/s & 79 km/s & 77 km/s \\
  &  $\gamma$ & $9.8\times 10^{-8}$ km$^{-1}$ & $3.4\times 10^{-8}$ km$^{-1}$ & $6.2\times 10^{-8}$ km$^{-1}$ \\
\hline
 In situ & $t_{sh}$ & 2012.1.22 05:32:58 UT & 2012.1.24 14:40:06 UT & 2012.3.8 10:30:45 UT \\
 observations & $V_{sh}$ & 466 km/s & 719 km/s & 1088 km/s \\
 at & $t_{ICME}$ & 2012.1.23 00:00:00 UT & 2012.1.25 12:00:00 UT & 2012.3.9 05:00:00 UT \\
 WIND & $V_{ICME}$ & 455 km/s & 602 km/s & 717 km/s \\
\hline
 \multirow{16}{*}{Outputs}& $t_{sh,HM}$ & 2012.1.22 03:11:00 UT & 2012.1.24 23:06:00 UT & 2012.3.8 17:55:00 UT \\
 & $V_{sh,HM}$ & 665 km/s & 900 km/s & 950 km/s \\
 & $\Delta TT_{sh,HM}$ & 2.37 hr & -8.43 hr & -7.40 hr \\
 & $\Delta V_{sh, HM}$ & -199 km/s & -181 km/s & 138 km/s \\
 & $t_{sh,DGSPM}$ & 2012.1.21 21:14:04 UT & 2012.1.25 01:13:15 UT & 2012.3.8 13:34:18 UT \\
 & $V_{sh,DGSPM}$ & 653 km/s & 799 km/s & 873 km/s \\
 & $\Delta TT_{sh,DGSPM}$ & 8.31 hr & -10.55 hr & -3.06 hr \\
 & $\Delta V_{sh,DGSPM}$ & -187 km/s & -80 km/s & 215 km/s \\
 & $t_{sh,DGSTOA}$ & 2012.1.21 21:53:12 UT & 2012.1.25 03:40:06 UT & 2012.3.8 12:12:36 UT \\
 & $V_{sh,DGSTOA}$ & 623 km/s & 718 km/s & 905 km/s \\
 & $\Delta TT_{sh,DGSTOA}$ & 7.66 hr & -13.0 hr & -1.70 hr \\
 & $\Delta V_{sh,DGSTOA}$ & -157 km/s & 1 km/s & 183 km/s \\
 & $t_{ICME,DBM}$ & 2012.1.23 05:10:02 UT & 2012.1.25 06:59:48 UT & 2012.3.9 17:24:36 UT \\
 & $V_{ICME,DBM}$ & 383 km/s & 604 km/s & 444 km/s \\
 & $\Delta TT_{ICME,DBM}$ & -5.17 hr & 5.0 hr & -12.41 hr \\
 & $\Delta V_{ICME,DBM}$ & 72 km/s & -2 km/s & 273 km/s \\
\hline
\end{tabular}
\begin{tabular}{l}
\textbf{Note.} $t_M$, and $R_M$ are the time and distance when CME
front reaches the maximum speed ($V_M$). $AW_{CME}$ is the CME \\
angular width observed by SOHO/LASCO. $V_{SW}$ is the solar wind
speed. $V_A$, $C_S$, and $V_F$ stand for the  Alfv\'en speed,\\
sound speed, and fast-mode wave speed, respectively. $\gamma$ is the
drag coefficient adopted by DBM. $t_{sh}$ and $V_{sh}$ are the \\
shock arrival time and local speed at WIND. $t_{ICME}$ and
$V_{ICME}$ denote the arrival time and local speed of the ICME at \\
WIND; they correspond to the sheath's back boundary for Case 2.
$t_{sh,HM}$ and $V_{sh,HM}$ represent the shock arrival time and \\
propagation speed at Earth predicted by HM; $\Delta TT_{sh,HM}$ and
$\Delta V_{sh,HM}$ are their errors. Similarly, $t_{sh,DGSPM}$, $V_{sh,DGSPM}$, \\
$\Delta TT_{sh,DGSPM}$,$\Delta V_{sh,DGSPM}$ represent the
prediction results and errors of DGSPM for the shock;
$t_{sh,DGSTOA}$, $V_{sh,DGSTOA}$, \\ $\Delta TT_{sh,DGSTOA}$,
$\Delta V_{sh,DGSTOA}$ represent those of DGSTOA for the shock;
$t_{sh,DBM}$, $V_{sh,DBM}$, $\Delta TT_{sh,DBM}$, $\Delta
V_{sh,DBM}$ represent\\ those of DBM for the ICME. Arrival times of
the shock and ICME at WIND and the HM triangulation results are
taken from \\ \citep{Liuetal2013}.
\end{tabular}
\end{table}
\clearpage

\begin{table}
\caption{The mean-absolute prediction errors for the CME/shock
arrival times and their local propagation speeds at Earth predicted
by the HM triangulation, DGSPM, DGSTOA, and DBM.} \centering
\begin{tabular}{ccccc}
\hline
\hline
 & Prediction & Error of  Arrival & Error of local &  \\
 &  Method   & ~ Time (hr) & ~ speed (km/s) &\\
\hline
  & HM triang.~~ & 6.07 & 173 &  \\
  & DGSPM ~~ & 7.31 & 161 &  \\
  & DGSTOA ~~ & 7.45 & 114 &  \\
  & DBM ~~ & 7.53  & 116 &  \\
\hline
\end{tabular}
\end{table}
\clearpage

\end{document}